\documentclass[conference]{IEEEtran}
\IEEEoverridecommandlockouts
\usepackage{cite}
\usepackage{amsmath,amssymb,amsfonts}
\usepackage{algorithmic}
\DeclareMathOperator{\diag}{diag}
\usepackage{graphicx}
\usepackage{textcomp}
\usepackage{xcolor}
\usepackage{array}
\usepackage{caption}
\usepackage{subcaption}
\newtheorem{remark}{\textbf{Remark}}
\makeatletter
\newcommand{\thickhline}{%
    \noalign {\ifnum 0=`}\fi \hrule height 1pt
    \futurelet \reserved@a \@xhline
}
\newcolumntype{"}{@{\hskip\tabcolsep\vrule width 1pt\hskip\tabcolsep}}
\makeatother
\def\BibTeX{{\rm B\kern-.05em{\sc i\kern-.025em b}\kern-.08em
    T\kern-.1667em\lower.7ex\hbox{E}\kern-.125emX}}
\begin{document}

\title{Extended NYUSIM-based MmWave Channel Model and Simulator for RIS-Assisted Systems}

\author{\IEEEauthorblockN{Aline~Habib\IEEEauthorrefmark{1}\IEEEauthorrefmark{2}, Israa~Khaled\IEEEauthorrefmark{1}, Ammar~El Falou\IEEEauthorrefmark{4}, Charlotte Langlais\IEEEauthorrefmark{1}}\\
\IEEEauthorblockA{\IEEEauthorrefmark{1} Mathematical and electrical engineering department, CNRS UMR 6285 Lab-STICC, IMT Atlantique, Brest, France}
\IEEEauthorblockA{\IEEEauthorrefmark{2} Faculty of Engineering, Lebanese University, Tripoli, Lebanon}
 \IEEEauthorblockA{ \IEEEauthorrefmark{4} CEMSE Division, King Abdullah University of Science and Technology (KAUST), Saudi Arabia}
Email: \{aline.habib, israa.khaled, charlotte.langlais\}@imt-atlantique.fr, ammar.falou@kaust.edu.sa} 

\maketitle
\begin{abstract}
    Spectrum scarcity has motivated the exploration of the millimeter-wave (mmWave) band as a key technology to cope with the ever-increasing data traffic. However, in this band, radiofrequency waves are highly susceptible to transmission loss and blockage. Recently, reconfigurable intelligent surfaces (RIS) have been proposed to transform the random nature of the propagation channel into a  programmable and controllable radio environment. This innovative technique can improve mmWave coverage. However, most works consider theoretical channel models. In order to fill the gap towards a realistic RIS channel simulator, we extend the 3D statistical channel simulator NYUSIM based on extensive measurements to help model RIS-assisted mmWave systems. We validate the extended simulator analytically and via simulations. In addition, we study the received power in different configurations. Finally, we highlight the effectiveness of using RIS when the direct link is partially blocked or non-existent.
\end{abstract}

\vspace{-0.1cm}
\section{Introduction}
To meet the high data rate requirements of 5G systems and beyond, millimeter-wave (mmWave) communications are emerging as an essential technology thanks to the enormous available bandwidth \cite{9350499}. However, mmWaves are highly vulnerable to oxygen absorption and rain attenuation and suffer from penetration loss.  All these limit the coverage of mmWave signals \cite{9568459}. On the other hand, a recent emerging hardware technology called reconfigurable intelligent surfaces (RIS) can improve transmission coverage while significantly reducing power consumption. A RIS is a thin, plane meta-surface composed of several low-cost passive reflecting elements connected to a controller that transfers commands from the base station (BS) to each RIS element \cite{9765815,9397266,habib2023reconfigurable}. The latter reflects the incident wave as a beam in the desired direction according to a controlled phase shift\cite{9593206}. The implementation of RIS deals with two  connections: transmitter (TX)-RIS link and RIS-receiver (RX) link. Thus, when the direct path is blocked or largely attenuated, the coverage area and connectivity may be increased thanks to RIS \cite{9201413}. 
This work considers RIS-assisted mmWave systems, in which we focus on modeling the corresponding channels to facilitate the performance evaluation of these systems. Most works on RIS-assisted systems  consider theoretical channel models such as Rician for line-of-sight (LoS) environments and Rayleigh when there are only non-LoS (NLoS) paths \cite{9739892,9738798,9606297,8982186}. Recently, the authors in \cite{9397266} aim to fill the gap towards realistic channel modeling and simulator, and they propose a novel physical channel simulator, called SimRIS. This simulator is based on statistical modeling and can be used in indoor and outdoor environments at 28 and 73 GHz frequencies. Specifically, SimRIS generates the channel coefficients in spatial and temporal domains using several statistical channel models \cite{9541182}. However, the coherence of the utilization of those different  channel parameters in one unique channel model is not verified by field measurements in mmWave environments. 

To this end, we leverage one well-known and realistic mmWave channel model and simulator, NYUSIM \cite{7501500}, and we extend it to handle the RIS as well. This channel can be applied in diverse environments, mono-path or multi-paths e.g., urban and rural areas. In addition, it is designed for either single-input single-output (SISO) or multiple-input multiple-output (MIMO) systems in single-user (SU) or multi-user (MU) scenarios. Specifically, it generates the channel coefficients between a TX and a/multiple RX. Besides, this geometric simulator provides the spatial distribution of RXs, i.e., separation distances from TX and their spatial directions. The proposed extended simulator, denoted as NYUSIM-RIS, can be adopted with SU/MU-SISO/MIMO systems. We derive the link budget for RIS-assisted SU-SISO systems using NYUSIM-RIS and validate its effectiveness analytically and via simulations. 

    

The rest of the paper is structured as follows. The system model is introduced in Section \ref{section2}. Section \ref{sec:NYUSIM-RIS} presents our proposed extended NYUSIM-based channel simulator for RIS-assisted MIMO systems. In Section \ref{section4}, we compare the link budget theoretically and with NYUSIM-RIS simulator. Numerical results are presented in Section \ref{section5}. Finally, Section \ref{conclusion} concludes the paper.

\textit{Notations}: Bold lower and upper case letters represent vectors
and matrices, respectively. $\textbf{H}^T$ denote the transpose of $\textbf{H}$. $\mathbb{C} ^{m\times n}$  denotes the space of $m\times n$
complex matrices. 
\section{System Model}\label{section2}
\begin{figure}
	\centering
	\includegraphics[width=1.25\columnwidth, trim=130 100 30 65, clip]{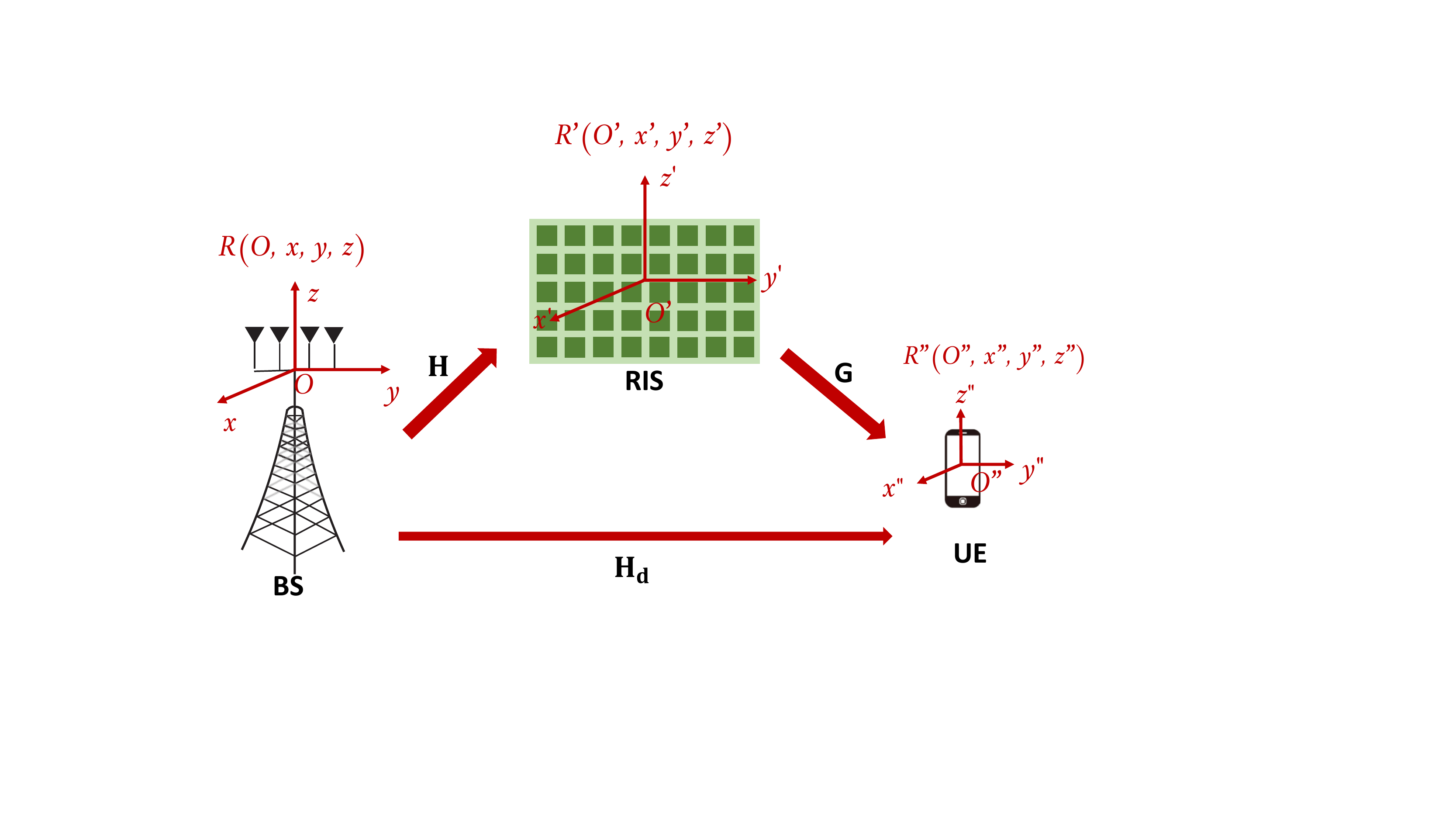}
	\caption{A Downlink RIS-assisted SU-MIMO transmission.}
	\label{fig:ris }
	\vspace{-0.3cm}
\end{figure}
We consider a narrowband  \footnote{Wideband systems can also be modeled using NYUSIM \cite{poddar_nyusim_nodate}.} mmWave  downlink SU-MIMO system assisted by a RIS, see Fig. \ref{fig:ris }.
Denote $M_{t}$ and $M_{r}$ as the number of antennas at the BS and at the user equipment (UE), respectively, and $N$  as the number of RIS  elements.
In the presence of RIS, three channels exist, 1)
$\textbf{H}$ $\in$ $ \mathbb{C} ^{N\times M_{t}}$: the BS-RIS channel, 2)  $\textbf{G}$ $\in$ $ \mathbb{C} ^{M_{r}\times N}$: the RIS-UE channel, and 3) $\textbf{H}_{d}$ $\in$ $ \mathbb{C} ^{M_{r}\times M_{t} }$: the direct BS-UE channel. These three MIMO channels can be obtained in the same way. For instance, the channel $\textbf{H}_\text{MIMO}\in \mathbb{C} ^{M_{2}\times M_{1}}$ between any TX and any RX with $M_1$  and $M_2$ antennas can be expressed as follows \cite{khaled2020jointSDMA}
\addtolength{\topmargin}{0.07in}
\begin{equation} 
\textbf{H}_\text{MIMO}=\sum_{n=1}^{N_p} \alpha_{n} e^{j\varphi_{n}}\textbf{a}_R(\vec{\Theta}_{n}^a)\textbf{a}_T^H(\vec{\Theta}_{n}^d),
	\label{eq:channel_h}
 \end{equation}
with $N_p$ the number of paths in the TX-RX channel, $ \varphi_{n} $ and $ \alpha_{n} $ the phase and the amplitude of the $ n $-th path. Note that $ \alpha_{n} $ includes the impact of distance, atmospheric attenuation, and foliage loss. $ \vec{\Theta}_{n}^d =(\theta_{n}^d,\phi_{n}^d)$, and  $ \vec{\Theta}_{n}^a =(\theta_{n}^a,\phi_{n}^a)$ the couple of azimuth and elevation angle-of-departure (AoD) and angle-of-arrival (AoA) of the $ n $-th path, respectively. Denote $\textbf{a}_T$ and $\textbf{a}_R$ as the array steering vectors  at the TX and the RX.  The LoS path\footnote{In NYUSIM, the spatial direction of each RX is represented by the AoD of the LoS path where this path is labeled by $1$.} has the highest power and is labeled by $ n = 1$. 
By definition, the steering vector $\textbf{a}(\vec{\Theta}) \in \mathbb{C} ^{M\times 1}$ of an array with  $M$ antennas is the set of phase-delays experienced by an incoming wave at each antenna to make a directive transmission at $\vec{\Theta}=(\theta,\phi)$ angle. 
We consider a uniform rectangular array (URA)\footnote{Using NYUSIM, we can use either URA or ULA at the transceiver.} located in the $ yOz $ plane as shown in Fig. \ref{fig:ris }, and having $ M_h $ horizontal and $ M_v $ vertical antennas along the $ y $-axis and the $z $-axis.  Thus, $M= M_h \times M_v $ and $\textbf{a}(\vec{\Theta}) \in \mathbb{C} ^{M\times 1}$ is given as follows \cite{khaled_multi-user_2021} 
\begin{equation}\label{eq:a_steering _vecor}\textbf{a}(\vec{\Theta}) \!\! = \!\! \left[\! 1, \! \cdots \! ,  e^{-j 2\pi\frac{d}{\lambda} \left\lbrace (M_h-1) \cos(\theta)\cos(\phi)+(M_v-1)\sin(\phi)\right\rbrace} \! \right ]^T
\end{equation} 
with $ j\triangleq\sqrt{-1} $ the imaginary unit, 
$\lambda$  the carrier wavelength and $d$  the inter-element spacing distance.
All the geometric parameters and channel coefficients are generated using NYUSIM \cite{7501500}. Specifically,
 this simulator adopts the time-cluster and spatial-lobe approach that couples temporal and spatial dimensions. For example, paths belonging to different temporal clusters can arrive at the same pointing angles.  
 This simulator also considers the mutual coupling and polarization of the antennas (co-polarization or cross-polarization)
\cite{zekri2022towards}. Note that there are two running modes for NYUSIM: static user, and moving user. In this work, we consider the static one.

\section{Extended NYUSIM-RIS Channel Model and Simulator}\label{sec:NYUSIM-RIS}
NYUSIM generates a realistic and statistical mmWave channel between a TX and a/multiple RX for SU/MU- SISO/MIMO systems. In addition, it provides the spatial distribution of RXs, i.e., separation distances from the TX and their spatial directions w.r.t. to the reference centered at the TX. However, for RIS-assisted SU-MIMO systems, there are three channels, $\textbf{H}$, $\textbf{G}$ and $\textbf{H}_{d}$, thus three references, namely, $\mathcal{R}(O, x, y, z)$, $\mathcal{R’}(O’, x’, y’, z’)$ and $\mathcal{R"}(O", x", y", z")$ centered at the BS, the RIS, and the UE as shown in Fig.\ref{fig:ris }. The RIS surface  coincides with the $y'O'z'$ plane. In our work, we leverage the design of NYUSIM for MIMO systems to also handle RIS-assisted MIMO systems. To do so, we apply the following steps that extend NYUSIM to NYUSIM-RIS.

\textit{{\textbf{Step 1: Generate the coefficients of the BS-RIS channel}}\label{step1}}\\
First, NYUSIM generates the BS-RIS channel matrix $\textbf{H}$ by considering the BS as the TX and the RIS as the RX. And $\textbf{H}$ can be expressed as follows
\begin{equation}
\textbf{H}=\sum_{n_1=1}^{N_{p1}} \breve{\alpha}_{n_1} e^{j\Breve{\varphi}_{n_1}}\textbf{a}_{RIS}(\vec{\breve{\Theta}}_{n_1}^a)\textbf{a}_{BS}^H(\vec{\breve{\Theta}}_{n_1}^d),
	\label{eq:channel_BS_IRS}
 \end{equation} 
with $N_{p1}$ the number of paths in the BS-RIS channel, $\Breve{\varphi}_{n_1} $ and $ \breve{\alpha}_{n_1} $ the phase and the amplitude of the $ n_1 $-th path, $ \vec{\breve{\Theta}}_{n_1}^d =(\breve{\theta}^d_{n_1},\breve{\phi}^d_{n_1})$ the couple of azimuth and elevation AoD relative to $\mathcal{R}$, and  $ \vec{\breve{\Theta}}_{n_1}^a =(\breve{\theta}^a_{n_1},\breve{\phi}^a_{n_1})$  the couple of azimuth and elevation AoA relative to  $\mathcal{R'}$ of the $ n_1 $-th path. Moreover, at the BS, an emitted power $P_{t}$ and a transmit antenna gain $G_t$ may be chosen. The receive antenna gain at RIS is fixed to 1.

Based on the BS-RIS separation distance, $d_1$, and the RIS direction w.r.t. $\mathcal{R}$ represented by $ \vec{\breve\Theta}_{1}^d =(\breve\theta_{1}^d,\breve\phi_{1}^d)$, the Cartesian coordinates of RIS in $\mathcal{R}$ are given by 
\begin{equation}\label{eq:omegaExp2}
\begin{cases}
x_\text{RIS}=\rho\sin(\breve{\phi}_{1}^d)\cos(\breve{\theta}_{1}^d), & \\
y_\text{RIS}=\rho\sin(\breve{\theta}_{1}^d)\sin(\breve{\phi}_{1}^d), & \\
z_\text{RIS}=\rho\cos(\breve{\phi}_{1}^d), & \\
\end{cases}
\end{equation}
with $\rho =d_1 / \sin(\breve{\phi}_{1}^d)$. Note that, throughout this step, we can fix the position of RIS in $\mathcal{R}$, i.e., $d_1$ and $ \vec{\breve\Theta}_{1}^d $, to generate \textbf{H}.

\textit{{\textbf{Step 2: Generate the coefficients in the RIS-UE channel}}}\\
Now we set NYUSIM such that RIS is the TX. The emitted power at the RIS $P_{t2}$ is equal to $\rho^2 P_{r1}$, where $\rho$ is the reflection coefficient at the RIS and $P_{r1}$ is the received power at the RIS obtained in the first step. We assume that there is no reflection loss (or amplification) and then $\rho$ =1. The channel $\textbf{G}$ between the RIS and the UE is expressed as follows 
\begin{equation}
\textbf{G}=\sum_{n_2=1}^{N_{p2}} \Tilde{\alpha}_{n_2} e^{j\Tilde\varphi_{n_2}}\textbf{a}_{UE}(\vec{\Tilde{\Theta}}_{n_2}^a)\textbf{a}_{RIS}^H(\vec{\Tilde{\Theta}}_{n_2}^d),
	\label{eq:channelIRS_user}
	\end{equation} 
	with $ N_{p2} $ the number of paths in the RIS-UE channel, $ \Tilde\varphi_{n_2} $ and $ \Tilde{\alpha}_{n_2} $ the phase and the amplitude of the $ n_2 $-th path, $ \vec{\Tilde{\Theta}}_{n_2}^d =(\Tilde{\theta}_{n_2},\Tilde{\phi}_{n_2})$ the couple of azimuth and elevation AoD relative to $\mathcal{R'}$, and 
 $ \vec{\Tilde{\Theta}}_{n2}^a =(\Tilde{\theta}_{n_2},\Tilde{\phi}_{n_2})$ the couple of azimuth and elevation AoA relative to $\mathcal{R"}$ of the $n_2 $-th path. The receive antenna gain at the UE $G_r$ can be set in NYUSIM.
	
Similarly to Step 1, based on the RIS-UE separation distance, $d_2$, and the UE direction w.r.t. $\mathcal{R'}$ represented by $ \vec{\Tilde{\Theta}}_{1}^d =(\Tilde{\theta}_{1}^d,\Tilde{\phi}_{1}^d)$, the Cartesian coordinates of the UE in $\mathcal{R'}$ are given by 
\begin{equation}\label{eq:omegaExp3}
\begin{cases}
x'_\text{UE}=\rho'\sin(\Tilde{\phi}_{1}^d)\cos(\Tilde{\theta}_{1}^d), & \\
y'_\text{UE}=\rho'\sin(\Tilde{\theta}_{1}^d)\sin(\Tilde{\phi}_{1}^d), & \\
z'_\text{UE}=\rho'\cos(\Tilde{\phi}_{1}^d), & 
\end{cases}
\end{equation}
with $\rho'=d_2/\sin(\Tilde{\phi}_{1}^d)$.

\textit{{\textbf{Step 3: Generate the coefficients of the direct BS-UE channel}}}\\
We still need the BS-UE channel, $\textbf{H}_{d}$. Since NYUSIM is a geometric simulator, we need to find first the geometric position of the UE relative to $\mathcal{R}$, i.e., the BS-UE separation distance, $d_\text{BS-UE}$, and the UE direction in $\mathcal{R}$, $ \vec{\Theta}_d =(\theta_d,\phi_d)$. This information is computed  according to Step 1 and Step 2 so that the resulting $\textbf{H}_{d}$ channel is coherent with $\textbf{H}$ and $\textbf{G}$. To find  $d_\text{BS-UE}$ and $ \vec{\Theta}_d =(\theta_d,\phi_d)$, we apply the following steps:
\begin{itemize}
    \item Calculation of Cartesian coordinates of the UE in $\mathcal{R}$ by applying the following transformation of references:
\begin{equation}\label{eq:eqRefe}
\begin{cases}
x_\text{UE}=x_\text{RIS} + x'_\text{UE},& \\
y_\text{UE}=y_\text{RIS} + y'_\text{UE}, & \\
z_\text {UE}= z_\text {RIS} + z'_\text {UE}. & \\
\end{cases}
\end{equation}
\item Calculation of spherical coordinates of the UE in $\mathcal{R}$:
\begin{equation}\label{eq:cartspher}
\begin{cases}
\rho_d=\sqrt{x_\text{UE}^2+y_\text{UE}^2+z_\text{UE}^2},& \\
\phi_d=\arccos{(z_\text{UE}/\rho)},& \\
\theta_d=\arctan{(y_\text{UE}/x_\text{UE})}. & \\
\end{cases}
\end{equation}
Here, the UE direction $\vec{\Theta}_d =(\theta_d,\phi_d)$  is obtained
\item Calculation of $d_\text{BS-UE}$ as follows:
\begin{equation}\label{eq:sphtocart}
d_\text{BS-UE}=\rho_d\sin(\phi_d),
\end{equation}
\end{itemize}
Once the geometric position of the UE in $\mathcal{R}$ is defined, we set NYUSIM to generate the coefficients of the direct link channel. 

These three steps implemented in NYUSIM-RIS for RIS-assisted SU-MIMO systems can also be applied with  RIS-assisted MU-MIMO systems. Specifically, with $K$ UEs, Step 1 remains the same, while Step 2 and Step 3 are repeated $K$ times. For the rest, we consider only RIS-assisted SU-SISO systems as a preliminary study to confirm the validity of NYUSIM-RIS. 

\section{Link Budget for RIS-assisted SISO Systems using NYUSIM-RIS }\label{section4}

In this section, 
we calculate the link budget for RIS-assisted SU-SISO systems using NYUSIM-RIS. And we compare it with that generated using the log-distance model in \cite{alfattani2021link}, denoted as Log-Dist-RIS. The authors in \cite{alfattani2021link} assume that the direct link between the TX and RX  is weak and they neglect it. For a fair comparison, we consider the same assumption. 
\subsection{Link budget of mmWave systems using NYUSIM }
\noindent Using NYUSIM, the path loss $P_L^{\text{NYU}}$ between  TX and  RX results from free space, shadowing, atmospheric attenuation, and foliage \cite{7501500}:
\begin{enumerate}
    \item The free space loss is defined as
    $P_L^{\text{FS}}(d_0)= 20\log\left(\frac{4\pi d_0}{\lambda}\right)$, with $d_0$ the reference distance.
    \item The path loss at distance  $d$ is defined as $P_L(d)= 10 \mu \log\left(\frac{d}{d_0}\right)$, with $\mu$ the path loss exponent depending on the environment type (e.g., $\mu \ge 3$ for urban environments) and $d$ is the TX-RX separation distance.  
    \item The shadowing loss is defined as  $P_L^{\text{Sh}}={s} \delta $, with $s$ the shadow factor standard deviation in dB and $\delta$ a normally distributed random number. 
    \item The path loss resulting from atmospheric attenuation is defined as $P_L^{\text{At}}=A_t d$. Denote $A_t$ as the attenuation factor which depends on temperature, partial pressures, water permittivity, water vapor, and rain attenuation. 
    \item The foliage loss is defined as $P_L^{\text{Fol}}=A_\text{Fol} d_\text{Fol}$. Denote $A_\text{Fol}$ as the foliage attenuation in $\text{dB/m}$ ranging from 0 to 10 dB, and $d_\text{Fol}$ as the distance within the foliage, which can be set to any non-negative value. 
\end{enumerate} 
Thus, 
\begin{equation}\label{PL_eq}
P_L^{\text{NYU}}=P_L^{\text{FS}}(d_0)+P_L(d)+P_L^{\text{Sh}}+P_L^{\text{At}}+P_L^{\text{Fol}}.
 \end{equation} 
 And the power $P_{\text{RX}}^{\text{NYU}}$ received at the RX is given by \cite{7501500}
\begin{equation}\label{Pr_eq_dbm}
P_{\text{RX}}^{\text{NYU}}[\text{dBm}]=P_{\text{TX}}-P_L^{\text{NYU}},
 \end{equation}  
 with $P_{\text{TX}}$ the power emitted by the TX.

\subsection{Link budget of RIS-assisted mmWave systems using NYUSIM-RIS }
\noindent  In \cite{alfattani2021link}, the authors   discuss two types of RIS-assisted communications, namely, 1) the specular reflection paradigm, and 2) the scattering reflection paradigm. In the first case, the RIS is considered in the near field and the path loss depends on the summation of the distances BS-RIS and RIS-UE.
In the second case, the RIS is considered in the far field and the path loss is based on the product of the BS-RIS and RIS-UE separation distances. In this paper,  we consider the scattering reflection paradigm. Moreover, to derive the link-budget analysis, we assume a mono-path model in the BS-RIS and RIS-UE channels, i.e., $N_{p1}=N_{p2}=1$ as in \cite{alfattani2021link}.
    
For a RIS-assisted SU-SISO system, the power received at the UE $P_{r}^\text{RIS}$  via RIS can be expressed as follows \cite{alfattani2021link}
\begin{equation}\label{eq:alphasquare}
P_{r}^\text{RIS}=\sum_{i=1}^{N}\beta_i^{2},
\end{equation}
with $\beta_i$ the channel coefficient  resulting from the two cascaded channels, i.e., BS-RIS and RIS-RX, for each RIS element, given by
\begin{equation}\label{eq:alpha}
\beta_i=\sqrt{P_tG_tG_r}h_{i}g_{i}\rho_ie^{-j\psi_i}, 
\end{equation}
where $h_{i}$ (resp. $g_{i}$) represents  the channel coefficient of the link between the $i$-th reflector and  the BS (resp. the UE). Besides, $\psi_i$ is the adjusted phase shift at the RIS and $\rho_i$ is the reflection coefficient of the $i$-th reflector, equal to 1 as mentioned in section \ref{sec:NYUSIM-RIS}. 
In  \cite{alfattani2021link}, the Log-Dist-RIS is applied to compute $h_{i}$ and $g_{i}$. However, to derive the link budget of NYUSIM-RIS, we replace $h_{i}$ and $g_{i}$ by their expressions according to Step 1 and Step 2 in Section \ref{sec:NYUSIM-RIS}. Therefore, with a single antenna at both the BS and the UE, (\ref{eq:channel_BS_IRS}) and (\ref{eq:channelIRS_user}) are simplified to 
 \begin{equation}
\textbf{h}=\Breve{\alpha}_{1} e^{j\Breve{\varphi}_{1}}\textbf{a}_\text{RIS}(\vec{\Breve{\Theta}}_{1}^a),
	\label{eq:channel_BS_IRS_SIMO}
 \end{equation} 
 \begin{equation} 
\textbf{g}=\Tilde{\alpha}_{1} e^{j\Tilde\varphi_{1}}\textbf{a}_\text{RIS}^H(\vec{\Tilde{\Theta}}_{1}^d).
\label{eq:channel_IRS_user_SIMO}
 \end{equation}
So, the $i$-th element $h_i$ and $g_i$ of $\textbf{h}$ and $\textbf{g}$ are respectively given by 
\begin{equation} \label{eq:hi}
 h_i=\Breve{\alpha}_{1} e^{j(\Breve{\varphi_{1}}+\Breve{z}_i)}, 
 \end{equation}
 \begin{equation}\label{eq:gi}
 g_i=\Tilde{\alpha}_{1} e^{j(\Tilde{\varphi_{1}} -\Tilde{z}_i)},
 \end{equation}
 with $\Breve\varphi_{1}$ and $\Tilde{\varphi_{1}}$ are the phase of the LoS path in the  BS-RIS and RIS-UE channels, $\Breve{z_i}$ and $\Tilde{z_i}$ are the phase shift resulting from $\textbf{a}_\text{RIS}(\vec{\Breve{\Theta}}_{1}^a)$ and $\textbf{a}_\text{RIS}(\vec{\Tilde{\Theta}}_{1}^d)$. $\Breve{\alpha}_{1}$ and $\Tilde{\alpha}_{1}$ are given according to NYUSIM as follows
 \begin{equation}\label{alpha1}
\Breve{\alpha}_{1}=\sqrt{\frac{P_{r1}}{P_{t}}},
 \end{equation}
 \begin{equation}\label{alpha2}
\Tilde{\alpha}_{1}=\sqrt{\frac{P_{r2}}{P_{t2}}},
 \end{equation}
with $P_{r2}$ the received power at the UE and the other parameters defined in Section \ref{sec:NYUSIM-RIS}.

For a fair comparison between Log-Dist-RIS \cite{alfattani2021link} and NYUSIM-RIS, we neglect the foliage, the shadowing, and the atmospheric losses. So, the total path loss $P_L$ in (\ref{PL_eq}) for a RX at a distance $d$ from the TX will be as
 \begin{equation} \label{pl}
P_L =20\log\left(\frac{4\pi d_0}{\lambda}\right)+10 \mu \log\left(\frac{d}{d_0}\right).
 \end{equation}
 Hence, $P_{\text{RX}}$ in (\ref{Pr_eq_dbm}) (in $W$) can be rewritten by
  \begin{equation}\label{eq:PR_w}
P^{\text{NYU}}_{\text{RX}}[W] =P_{\text{TX}}\left(\frac{\lambda} {4\pi d_0}\right)^2\left(\frac{d_o}{d}\right)^\mu.
 \end{equation} 
 Similarly, we can obtain $P_{r1}$ and $P_{r2}$ as follows
\begin{equation}\label{eq:PR_1}
P_{r1} =P_{t} \left(\frac{\lambda} {4\pi d_0}\right)^2\left(\frac{d_o}{d_1}\right)^\mu,
 \end{equation}
\begin{equation}\label{eq:PR_2}
P_{r2} =P_{t2} \left(\frac{\lambda} {4\pi d_0}\right)^2\left(\frac{d_o}{d_2}\right)^\mu,
 \end{equation}
 where $d_1$ and $d_2$ are the BS-RIS and RIS-UE separation distances, respectively.
 
 According to (\ref{eq:PR_1}) and (\ref{eq:PR_2}), $\Breve{\alpha}_{1}$ in (\ref{alpha1}) and $\Tilde{\alpha}_{1}$ in (\ref{alpha2}) will be  as follows
  \begin{equation}\label{alpha11}
\Breve{\alpha}_{1} =\sqrt{\left(\frac{\lambda} {4\pi d_0}\right)^2\left(\frac{d_o}{d_1}\right)^\mu},
 \end{equation} 
  \begin{equation}\label{alpha22}
\Tilde{\alpha}_{1} =\sqrt{\left(\frac{\lambda} {4\pi d_0}\right)^2\left(\frac{d_o}{d_2}\right)^\mu}.
 \end{equation}

 To maximize $P_{r}^\text{RIS}$ in (\ref{eq:alphasquare}), the optimal phase shift $\psi_i^{\text{Opt}}$ of the $i^{th}$ reflector must compensate the phase shift coming from  $h_i$  and  $g_i$, as seen in (\ref{eq:alpha}). Thus, according to (\ref{eq:hi}) and (\ref{eq:gi}), $\psi_i^{\text{Opt}}$ can be expressed as follows
\begin{equation}\label{eq:angle_opt}
\psi_i^{\text{Opt}}=\Breve{\varphi}+\Breve{z}_i+\Tilde{\varphi} -\Tilde{z}_i. 
\end{equation}
By replacing $\Breve{\alpha}_{1}$ in (\ref{eq:hi}) and $\Tilde{\alpha}_{1}$ in (\ref{eq:gi}) by their values in (\ref{alpha11}) and (\ref{alpha22}), the optimal power $P_r^{\text{RIS-Opt}}$ received when the RIS applies the optimal phase shift, based on (\ref{eq:alphasquare}), can be expressed as follows
 \begin{equation}\label{eq:pN}
P_{r}^{\text{RIS-Opt}}=P_tG_tG_r\left(\frac{\lambda}{4\pi}\right) ^4\frac{(d_0)^{2\mu-4}}{{(d_1d_2)^\mu}}N^2.
\end{equation}

Thus, we obtain the same expression (\ref{eq:pN}) for the received power as (14) in \cite{alfattani2021link}, based on  the channels generated by NYUSIM.  However, unlike Log-Dist-RIS \cite{alfattani2021link}, NYUSIM-RIS can consider several environmental effects, such as temperature, rainfall, foliage, shadowing, etc. For instance, if we take the shadowing into account, the received power at the UE $P_{r}^{\text{RIS-Opt-Sh}}$  can be written as follows
\begin{equation}\label{eq:pNN}
P_{r}^{\text{RIS-Opt-Sh}}=P_tG_tG_r\left(\frac{\lambda}{4\pi }\right)^4\frac{(d_0)^{2\mu-4}}{{(d_1d_2)^\mu}}10^{\frac{-s \delta}{5}}N^2.
\end{equation}

According to (\ref{eq:pN}), $P_{r}^{\text{RIS-Opt}}$ is maximized when the RIS is close to either TX or RX due to the product of distances in the denominator. This will be verified in Section \ref{distances}.
Next, both (\ref{eq:pN}) and (\ref{eq:pNN}) will be analyzed via simulations.

\section{ Numerical Results}\label{section5}
\begin{table}

	\centering
	\normalsize
	\caption{Simulation Parameters.}
	\begin{tabular}{|c|c|}
		\hline
		\textbf{Parameters} & \textbf{Value}\\
		\hline
		Carrier frequency & 30 [GHz]\\
		\hline
		Channel bandwidth & 20 [MHz]\\
		\hline
		RIS elements separation distance & $ \lambda/2 $\\
		\hline
		Transmission power at the BS, $P_t$ & 30 [dBm]\\
	
		\hline
       BS gain, $G_t$  & 0 [dBi]\\
       \hline
        UE gain, $G_r$ & 0 [dBi]\\
        \hline
	\end{tabular} 
	\label{tab:Parametres}
\end{table}

In this section, we verify the feasibility of the extended NYUSIM-RIS simulator for RIS-assisted mmWave SISO systems through numerical simulations and investigate the impact of some system parameters on the overall performance. Table \ref{tab:Parametres} summarizes the adopted simulation parameters, similar to those in \cite{alfattani2021link}. We set the positions of the BS and the RIS at (0,0,10) and (-50,50,10), respectively, similar to  \cite{9739892}.

\subsection{Number of RIS elements}\label{sec:simRes_nbRISelts}

\begin{figure}[t]
 \vspace{-0.7cm}
	\centering
	\includegraphics[width=1.3\columnwidth, trim=75 234 0 220, clip]{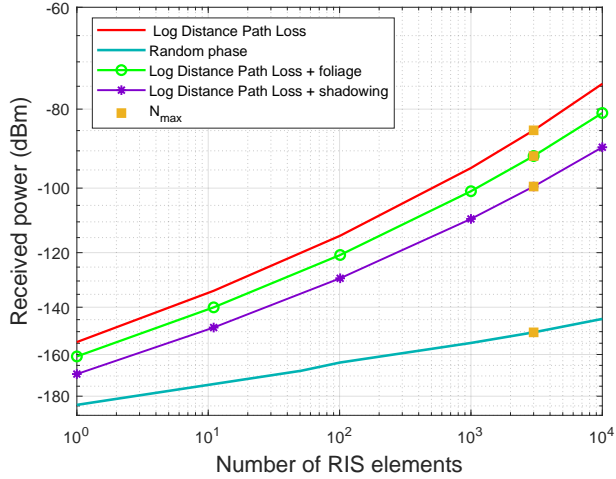}
	\caption{Received power  versus the number of RIS elements in different radio environments ( $d_1= 70.71$ m; $d_2 = 15 $ m).}  
	\label{fig:Prshad}
	\vspace{-0.2cm}
\end{figure}

Fig.~\ref{fig:Prshad} plots the received power  w.r.t. the number of RIS elements $N$, by considering three types of losses, i) log distance path loss $P_{r}^{\text{RIS-Opt}}$, according to (\ref{eq:pN}), ii) with shadowing $P_{r}^{\text{RIS-Opt-Sh}}$ for a constant value of $\delta$ = 1.8 dBm, according to (\ref{eq:pNN}), and iii) with the foliage loss. We consider only one realization of shadowing and foliage losses since the positions of BS, RIS, and UE are fixed in this work. Moreover, we also represent the received power based on random phase shifts at the RIS and the channel coefficients computed by our proposed extended simulator described in section \ref{sec:NYUSIM-RIS}.
  
From Fig.~\ref{fig:Prshad}, it is observed that $P_{r}^{\text{RIS-Opt}}$ increases with $N$, which is obviously seen  in (\ref{eq:pN}), where $P_{r}^{\text{RIS-Opt}}$ is proportional to  $N^2$. Indeed, on the one hand, the more elements are  implemented at the RIS, the more the path loss resulting from the cascaded channels is compensated. On the other hand, foliage loss and shadowing have a negative impact on the received power. It is also observed that the choice of random phase shifts degrades the received power and there is a clear need for optimization.

\begin{remark}
To verify the RIS scattering reflection paradigm, the distances $d_{1}$ and $d_{2}$ should be in the far-field region. Thus, $d_{1}>d_{\text{min}}$ and $d_{2}>d_{\text{min}}$  where $d_{\text{min}}$ is the  Rayleigh distance defined by $d_{\text{min}}$=$\frac{2S}{\lambda}$ with $S=\frac{\lambda^2}{4}N$ the RIS area  \cite{alfattani2021link}. Therefore, for fixed values of $d_1$ and $d_2$, the maximum number of RIS elements, $ N_{\max}$, that satisfy the far-field assumption is defined as
 \begin{equation}
   N_{\max}=\frac{2\min(d_1,d_2) }{\lambda}
\end{equation}
In our case, with $d_1= 70.71$ m and $d_2 = 15$ m, $N_{\max}$=3000.

\end{remark}

\subsection{Separation distances} \label{distances}

\begin{figure}[t]
 \vspace{-0.7cm}
	\centering
	\includegraphics[width=1.25\columnwidth, trim=75 234 0 220, clip]{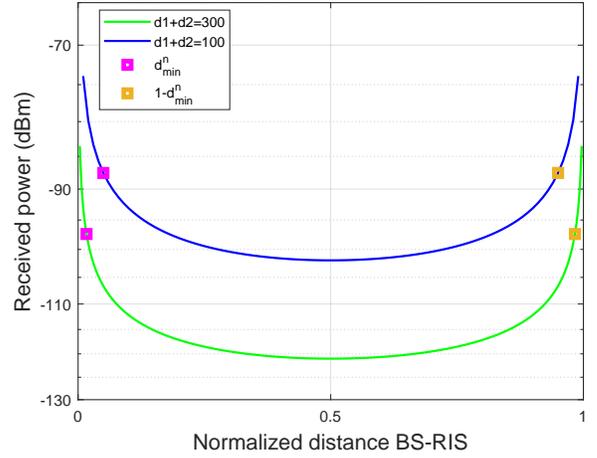}
	\caption{ Impact of the BS-RIS distance, $d_1$, and the RIS-UE distance, $d_2$, on the received power, $P_{r}^{\text{RIS-Opt}}$, for $N=1000$ and $d^{\text{n}}_{\min}$ the  Rayleigh distance normalized by $d_1+d_2$.}  
	\label{fig:pr_dist }
	\vspace{-0.2cm}
\end{figure}
 
To study the impact of the RIS position on the system performance, in Fig. \ref{fig:pr_dist }, we plot  the received power, $P_{r}^{\text{RIS-Opt}}$, versus the BS-RIS distance normalized by $d_1+d_2$, for $N=1000$.
This figure's symmetrical shape shows that the performance of the RIS-assisted system is the same if the RIS is close to the BS or the UE. Otherwise, with RIS in the middle,  the received power  has the lowest value. Moreover, we observe the detrimental path loss effect on the received power when the separation distance $d_1+d_2$ increases. Note that
$d_i \in (d_{\min},d_1+d_2-d_{\min}), i=\{1,2\}$ to satisfy the conditions of the RIS scattering reflection paradigm seen in Remark 1.
\subsection{Blockage on the direct link}\label{obstacless}

\begin{figure}
	\centering
         \vspace{-0.7cm}
	\includegraphics[width=1.25\columnwidth, trim=75 234 0 220, clip]{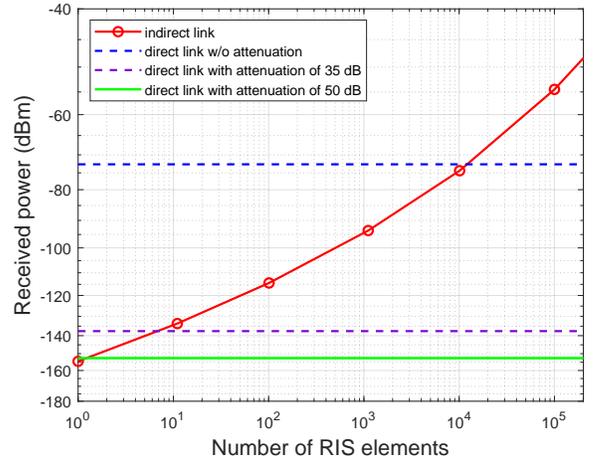}
	\caption{Received power versus the number of RIS elements for the direct link (distance from BS to UE equal to $105$ m) with and w/o obstacles and the indirect link via RIS ($d_1= 70.71$ m; $d_2 = 15$ m). }  
   \label{direct_indirect_Pr }
   \vspace{-0.35cm}
\end{figure}

Fig.  \ref{direct_indirect_Pr } plots the power received at the UE via the direct link, i.e., the BS-UE link, or the indirect one through the RIS, versus the number of RIS elements. 
Attenuation denoted by $\eta [dB] \in\{35, 50 \}$ \cite{7156092} represents different levels of blockage due to obstacles in the direct link. 
On the one hand, when there is no obstacle,  the received power via the direct link is  higher than  the received power via RIS for any number of RIS elements $\leq$ 10000.  On the other hand, when the direct link is attenuated with $50$ dB, the received power via the indirect link is  always higher. Therefore, the RIS is primarily needed in the case of a strongly attenuated direct link. And the combination of both direct and indirect links is all the more advantageous as the direct link is attenuated

\subsection{Multipath scenario rural and urban areas} 
\begin{figure}[t]
   \vspace{-0.7cm}
	\centering
	\includegraphics[width=1.25\columnwidth, trim=75 234 0 220, clip]{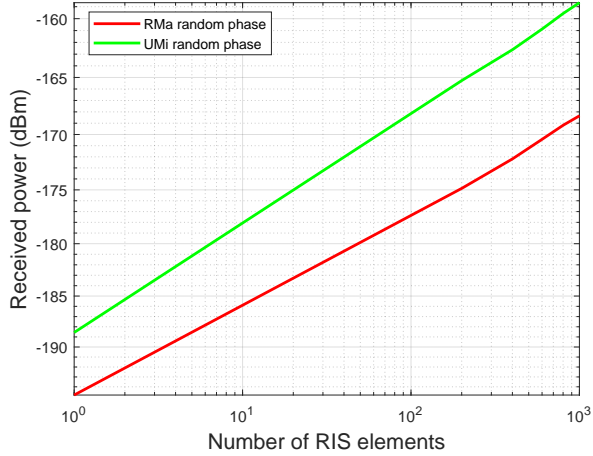}
	\caption{Received power  versus the number of RIS elements in the rural and urban environments.}  
	\label{fig:4}
	\vspace{-0.35cm}
\end{figure}
In the previous sections, we considered a mono-path environment with only LOS. 
In this section, we study the multipath environment for two realistic scenarios: Rural Macrocell (RMa) and Urban Microcell (UMi) areas \cite{7501500}. RMa involves one or two paths, whereas UMi is a multipath scenario, with 1 to 30 paths. Fig. \ref{fig:4} gives the received power in these 2 scenarios as a function of the number of RIS elements. The received power is given  as 
\begin{equation}\label{multPr}
    P_{r}=\lvert (\textbf{G}  \Psi\textbf{H})\rvert^2 
\end{equation}
where $\textbf{H}$ and $\textbf{G}$ are defined in section \ref{section2} and  $\Psi=\diag(e^{j\psi_1}, e^{j\psi_2},...e^{j\psi_N})$ is the RIS phase shift matrix. As an illustration,  $\psi_i$ follows a uniform distribution between 0 and $\pi$. 
 We observe that the RIS leverages the multipath diversity of UMi, even with random phase shifts, as the received power in the UMi case is much higher than in the RMa case. 

\section{Conclusion}\label{conclusion}
In this paper, we proposed an extension of the NYUSIM simulator to deal with RIS-assisted mmWave systems in realistic propagation environments. We derived the link budget of the proposed extended NYUSIM-RIS-based channel model in the far-field region, and we compared it to that of the log-distance-based channel in \cite{alfattani2021link}. However, to relieve the constraint of the limited number of RIS elements due to the far-field assumption, the near-field behavior of RIS and its performance have to be studied and are left as future work. Simulation results provide some well-known characteristics of RIS-assisted SISO systems, and we observed the multipath diversity in urban areas. In future work, it would be interesting to study the RIS-assisted MU-MIMO systems using the extended NYUSIM-RIS channel simulator in multipath environments. 
 

\bibliographystyle{IEEEtran}
\bibliography{Bibliography}


\end{document}